# Multi-GMR sensors controlled by additive dipolar coupling

J. Torrejon*, A. Solignac, C. Chopin, J. Moulin, A. Doll, E. Paul, C. Fermon and M. Pannetier-Lecoeur**.

*SPEC, CEA, CNRS, Université Paris-Saclay, CEA Saclay, 91191 Gif sur Yvette Cedex France*

## Abstract

Vertical packaging of multiple Giant Magnetoresistance (multi-GMR) stacks is a very interesting noise reduction strategy for local magnetic sensor measurements, which has not been reported experimentally so far. Here, we have fabricated multi-GMR sensors (up to 12 repetitions) keeping good GMR ratio, linearity and low roughness. From magnetotransport measurements, two different resistance responses have been observed with a crossover around 5 GMR repetitions: step-like (N<5) and linear (N≥5) behavior, respectively. With the help of micromagnetic simulations, we have analyzed in detail the two main magnetic mechanisms: the Neel coupling distribution induced by the roughness propagation and the additive dipolar coupling between the N free layers.

Furthermore we have correlated the dipolar coupling mechanism, controlled by the number of GMRs (N) and lateral dimensions (width), to the sensor performance (sensitivity, noise and detectivity) in good agreement with analytical theory. The noise roughly decreases in multi-GMRs as $\frac{1}{\sqrt{N}}$ in both regimes (low frequency 1/f and thermal noise). The sensitivity is even stronger reduced, scaling as $\frac{1}{N}$, in the strong dipolar regime (narrow devices) while converges to a constant value in the weak dipolar regime (wide devices). Very interestingly, they are more robust against undesirable RTN noise than single GMRs at high voltages and the linearity can be extended towards much larger magnetic field range without dealing with the size and the reduction of GMR ratio. Finally, we have identified the optimal conditions for which multi-GMRs exhibit lower magnetic field detectivity than single GMRs: wide devices operating in the thermal regime where much higher voltage can be applied without generating remarkable magnetic noise. These results open the path towards spintronics sensors connected and coupled in 3D with reduced noise, compact footprint, and mainly tuned by the dipolar coupling.

*\* Now at Institut Curie, PSL Research University, CNRS UMR, INSERM, Orsay, France.*
*\*\* Email: myriam.pannetier-lecoeur@cea.fr*

## I. INTRODUCTION

Spintronic devices based on Giant-Magnetoresistance (GMR)[1,2] exhibit large magnetoresistance variation which have been extensively used to read the magnetic information contained in hard disks during the last decades. Over the last years, GMR devices have also opened the market to weak magnetic sensing applications[3], mainly in automotive[4] or biological[5,6] systems, thanks to the remarkable enhancement of the their quality in terms of signal to noise ratio and CMOS compatibility, allowing small size integrated devices. Nevertheless, the strong correlation between magnetic field detection and sensor size is nowadays the main barrier to develop higher performance sensors at micro and nanoscale. Moreover, the detectivity in these devices is often limited by the presence of 1/f low frequency magnetic noise or random telegraphic noise (RTN) due to domain fluctuations in the sensing magnetic layer[7–9]. Several noise reduction schemes have been explored to mitigate these constrains ranging from sensors coupled to flux concentrators[10], MR sensors connected in series[11] or applying a pinning in the sensing layer[9,12–16]. But very often, such schemes have the disadvantage of increasing the footprint device and the resistance (and consequently thermal noise) and reducing the spatial resolution, which is very critical for local measurements.

An innovative strategy to reduce the noise, without dealing with the lateral size, is to connect vertically GMRs which combine low frequency noise and low resistance while maintaining a compact footprint. Vertical packaging systems, based on several spin valves stacked on top of each other and separated by a thick insulator spacer, have been theoretically proposed[17] but never experimentally demonstrated. One of the main challenges is related to the thick insulator spacer ($SiO_2$, $AlO_x$, etc) which should minimize the magnetostatic coupling while maintaining the roughness low.

In this work, we have fabricated multi-GMRs stack (up to 12 repetitions) separated by thin Ta spacer with low roughness, analyzed the main reversal magnetization mechanisms (Dipolar vs Neel couplings), successfully demonstrated noise reduction (in the 1/f and thermal regimes, respectively) and finally identified the optimal conditions towards lower magnetic field detectivity.

The paper is organized as follows. In Section II, we first introduce the experimental methods: sensor fabrication (sputtering, optical lithography and Ar ion etching), magnetotransport and noise set-ups and micromagnetic simulations details. We then present in Sections III and IV the magnetoresistance study in unpatterned multi-GMR thin films and yoke-shaped sensors, respectively, quantifying the role of Neel and dipolar couplings with the help of micromagnetic simulations. In Section IV, finally we evaluate the multi-GMR sensor performance parameters (sensitivity, noise and detectivity) as a function of number of GMRs, width and input voltage and the correlation with the dipolar coupling mechanism.

## II. METHODS

### A. Experimental details

Multi-GMR stacks have been deposited by sputtering on thermally oxidized (500 nm $SiO_2$) silicon wafers. Stacks comprise a generic top-pinned Synthetic AntiFerr-magnetic (SAF) spin-valve sequence {$Ni_{89}Fe_{19}$(5)/$Co_{90}Fe_{10}$ (2.1)/Cu(2.9)/$Co_{90}Fe_{10}$(2.1)/Ru(0.85)/$Co_{90}Fe_{10}$(2)/IrMn (7.5)} repeated N times, Figure 1(a), with N being varied from 1 to 12. Thicknesses are given in nm. The bottom free layer is composed by the NiFe/CoFe bilayer while the top pinned layer is a synthetic antiferromagnet CoFe/Ru/CoFe/IrMn. The multi-GMR stack is deposited onto a 3 nm Ta seed layer and ends with a (Ru (0.4)/ Ta(5)) capping bilayer. For N >1, each Top spin-valve sequence is separated from the next one by a 3 nm thick Ta layer, which allows dipolar coupling between the free layers while keeping roughness low. After deposition, stacks are annealed under vacuum for one hour at 473 K in 1T field applied in plane to set the pinned layers magnetization. Two thin film sets with nominally same composition were deposited at different dates: first film set A for the preliminary optimization of the multi-GMR stack configuration (see more details in Supplementary Figure S1) and then film set B for optimized and patterned multi-GMR sensors. Note that both exhibit the same GMR ratio and hysteresis but slightly different horizontal offset: offset in film set B is 0.3-0.5 mT larger than in film set A (See right panel Figure 1(c) for direct comparison).

The stacks are patterned combining optical lithography and Ar-ion etching in to yoke-shaped devices[18,19], Figure 2(a), which favors magnetic domain stabilization inside the main arm of the yoke[20]. Sensor widths varies from 1 μm to 30 μm, with a constant aspect ratio of 50:1 (L= 50 w) which leads to a nominal single element resistance $R_1$ of 750-800 Ω. Each repetition will therefore theoretically lead to $R_N=R_1/N$. Finally, the yokes are connected in a current-in-plane (CIP) configuration by Ta (5)/Cu (150)/Ta (5) contacts and passivated by a protective 150 nm thick $Al_2O_3$ layer deposited by sputtering as a last step in the microfabrication process.

Magnetotransport measurements (R-H curves) are performed by measuring the dc output voltage using 4 probes for the unpatterned films and 2 probes for the yokes in a home-made set-up. The external field is varied along the pinned layer direction (i.e. 90° from the yoke arm length) through a Helmholtz coil.

Noise measurements are performed in a magnetically shielded room through power spectrum recordings over the frequency range studied here (1Hz-3kHz) [19]. The GMR sensor is biased using a battery through a balanced Wheatstone bridge. The bridge output is amplified by an INA103 low-pass amplifier before a second step of amplification and band-pass filtering. An acquisition card acquires the temporal signal and a Fast Fourier Transform (FFT) is used to measure the noise spectral density. An AC field signal created by a coil ($7\mu T_{rms}$, 30 Hz) and applied along the sensitivity axis serves as calibration reference allowing to extract the limit of detection in nT.

## B. Micromagnetic modeling details

The magnetoresistance response of the multi-GMR sensors in Figure 2(b) are modeled by OOMMF micromagnetic simulations[21]. The geometry of the free layer is a rectangular prism with 7 nm thick, 4 μm width and 200 μm length and the gap between consecutive free layers is 28 nm. We assume that NiFe (5 nm thick) and CoFe (2 nm thick) layers are perfectly coupled by direct exchange with the following averaged magnetic parameters: saturation magnetization $M_s$= 850 kA/m and exchange constant $J_s$=10 pJ/m, weak uniaxial magnetic anisotropy $K_u$=200 J/m$^3$. The parameter values are in good agreement with literature[22]. The roughness of each free layer induces a Neel coupling (magnetostatic coupling between free layer i and its closest SAF layer which exhibits correlated roughness) which has been included in the simulations as bias field $H_{ni}$ (along the SAF pinned layer direction fixed during the annealing process). We have selected values close to experimental curves in figure 1: $H_{n1}$=0.8 mT, $H_{n2}$=1.3 mT, $H_{n3}$=1.4 mT, $H_{n4}$=1.5 mT and $H_{n6-8}$=1.6 mT). Indeed we consider for simplicity that the growth differences that could appear between the free layers (roughness, slight magnetization or thickness changes) are summarized in the Neel coupling. Very small magnetic field misalignment (1 deg) is applied in order to break the symmetry between clock-wise or anti clock-wise magnetization rotation.

To summarize, the stable magnetic configuration is given by the competition between the Zeeman energy, the magnetocrystalline and the shape anisotropies, the dipolar and the Neel interlayer couplings.

## III. Magnetoresistance of multi-GMRs in thin films: The role of Neel coupling variation

First, we have characterized and optimized multi-GMR thin films (up to N=12) from magnetoresistance measurements, Figures 1(b-c). During the optimization process, several multi-GMR combinations have been explored: multi-GMRs based on top-SAF spin valves, bottom-SAF spin valves, combination of both, different spacers (Ta and MgO), etc (see more details in Supplementary Figure S1). We have identified the best multi-GMR in terms of high GMR ratio (6-7 %), good linearity and low offset (≤ 3 mT for maximum N=12) for a stack composed by N repetitions of a Top-SAF spin valves where each single GMR unit (NiFe/CoFe/Cu/CoFe/Ru/CoFe/IrMn) is separated by 3 nm of Ta layer as shown figure 1(a). As N increases, the GMR ratio (measured at maximum field, ±25 mT) decreases very slightly from 7% (N=1) to 6 % (N=12), left panel in Figure 1(c). This is in contrast to single GMRs where an increase of magnetic volume results in a stronger decrease of the MR ratio. For example, for a single GMR with 15 nm thick NiFe (equivalent roughly to a multi-GMR with N slightly larger than 2), the GMR ratio decays down to 4% and the linearity is strongly degraded (Supplementary Figure S2).

On another hand, the average horizontal offset increases ~0.4 mT from the single to the double GMR stack but then it remains almost constant until N=8 and the offset increases again up to ~3 mT for the thickest stack with N=12, right panel in figure 1(c). The horizontal offset in a single GMR is mainly determined by the competition between oscillatory interlayer exchange coupling RKKY[23,24] and Neel[25,26] coupling. For optimal sensor performance, the RKKY interaction, which favors antiparallel (AP) configuration between free and pinned layers, should be likely compensated by Neel coupling which favors parallel (P) configuration. The oscillatory RKKY is mainly controlled by the Cu spacer thickness between pinned and free layers while the Neel coupling depends on the correlated roughness between the free and pinned layers setting up by dipoles at the homologous protrusions and bumps at the interfaces. For a single GMR, figure 1(b), we have obtained a good offset compensation (between 0.3 and 0.8 mT, for film set A and B, respectively) for a Cu thickness of 2.9 nm and 3 nm Ta as a buffer layer, which promotes the lowest roughness GMR stacks. For multi-GMR stacks, the thickness of the Cu layers is kept constant such that the offset variation is only ascribed to the Neel coupling variation as a consequence of roughness propagation between successive GMRs. The roughness degradation in multi-GMRs is only remarkable between the 1$^{st}$ and 2$^{nd}$ free layers and for multi-GMRs with N=12 where the linearity is also deteriorated. Note that another multi-GMR system based on inverted bottom-SAF spin valves (IrMn/CoFe/Ru/CoFe/Cu/CoFe/NiFe) exhibits 20-30% larger GMR ratio but its offset (roughness) and coercivity are strongly increased up to 5 times compared to top-SAF spin valves, Figure 1(c).

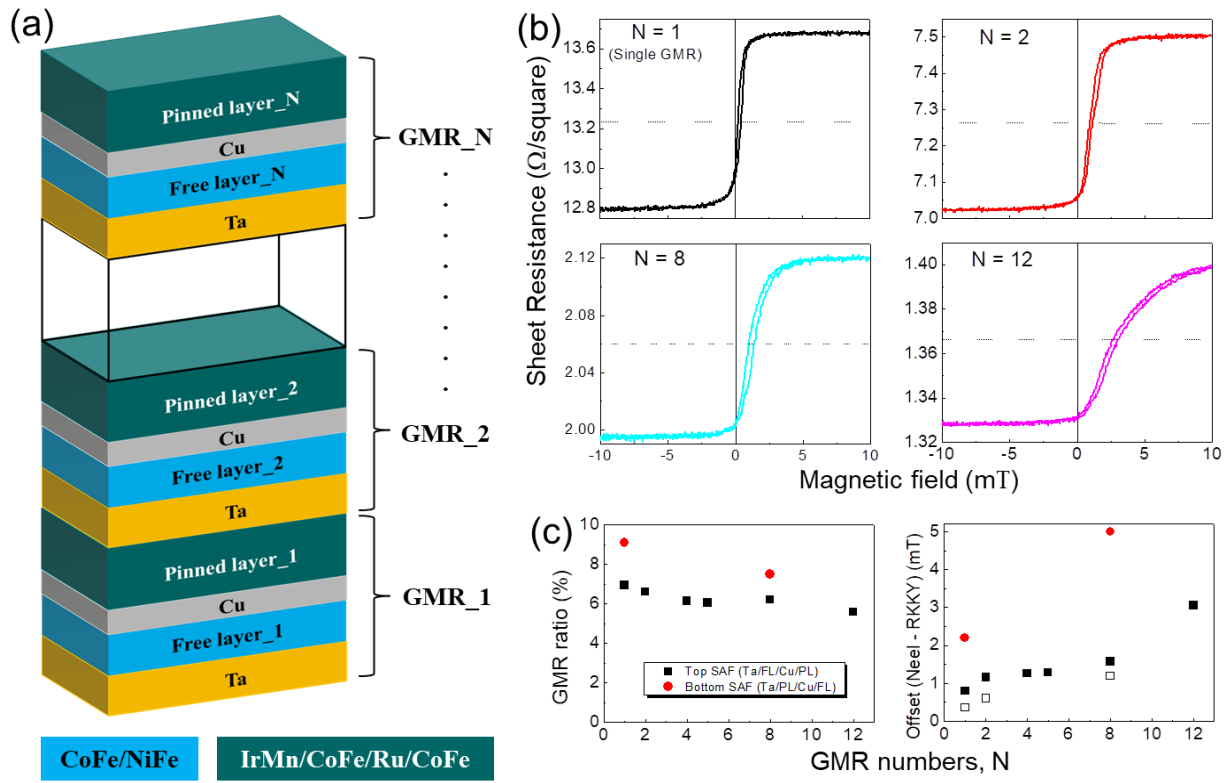

*Figure 1. Magnetoresistance in multi-GMR thin films. (a) Scheme of multi-GMR thin films based on several repetitions of top-SAF spin valves. (b) RH curves for 1, 2, 8 and 12 GMRs. (c) GMR ratio and offset evolution with the GMR number. Solid (Open) black squares corresponds to film set B (A).*

In general, the increase of the Neel field variation (roughness propagation) in each successive free layer degrades the linear behavior. However optimized top-SAF multi-GMRs in this study, figures 1(b), exhibit good linear response because the Neel coupling variation is small compared to the linear magnetic field range for each individual free layer, which is mainly governed by the magnetocrystalline anisotropy $H_{MC}$. And here we assume that $H_{MC}$ does not change between successive free layers leading to very similar resistance slopes. For non-optimized multi-GMRs, the Neel coupling variation is much larger inducing step-like behavior, Fig. S1(a). Even, very high roughness levels can vary the magnetocrystalline anisotropy in upper free layers degrading drastically the MR curve.

## IV. Magnetoresistance of yoke-shaped multi-GMR sensors: The role of additive dipolar coupling.

In the next section, we have firstly analyzed the magnetoresistance of yoke-shaped multi-GMR sensors as a function of number of GMR repetitions. Then we have performed micromagnetic simulations to understand and quantify the role of the additive dipolar coupling (magnetostatic fields interactions between successive free layers) over the Neel coupling variation (magnetostatic interaction between free and pinned layers with correlated roughness in each individual GMR).

### A. Experiments

A linear variation of the GMR resistance with the magnetic field with a low hysteresis is obtained when the easy magnetization axis of the free layer is set perpendicular to the pinned layer. This is achieved through the shape anisotropy[27] by patterning the yoke arm length perpendicular to the pinned layer magnetization as shown in Figure 2(a). The sensor magnetoresistance evolution with the number of GMR repetitions and constant width w= 4 µm is shown in Figures 2(b-c). We can clearly identified two different regimes. For N<5, the resistance shows a step-like response according to the number of repetitions. For N≥5, the RH curve exhibits a linear response. A similar trend has been observed when keeping N constant and varying the width (See Supplementary Figure S3). For N=2 (N=4), step-like behaviour is observed for w > 1 µm (w > 2 µm) and linear behaviour for w ≤ 1 µm (w ≤ 2 µm). Therefore, multi-GMR devices with large number of GMR and/or narrow sizes promote linear behaviour, which is more desirable for sensor applications.

These results indicate a strong effect of the dipolar couplings between neighbouring GMRs. In multi-GMR sensors, the dipolar fields are mainly governed by the free layers rather than the pinned layers. The pinned layer is a compensated SAF (with an effective thickness $t_{SAF}$~0.1nm) and its dipolar field is typically more than one order of magnitude smaller than the free layer ($t_{FL}$=7 nm). For this reason, we have only considered in our analysis the dipolar fields from the N free layers.

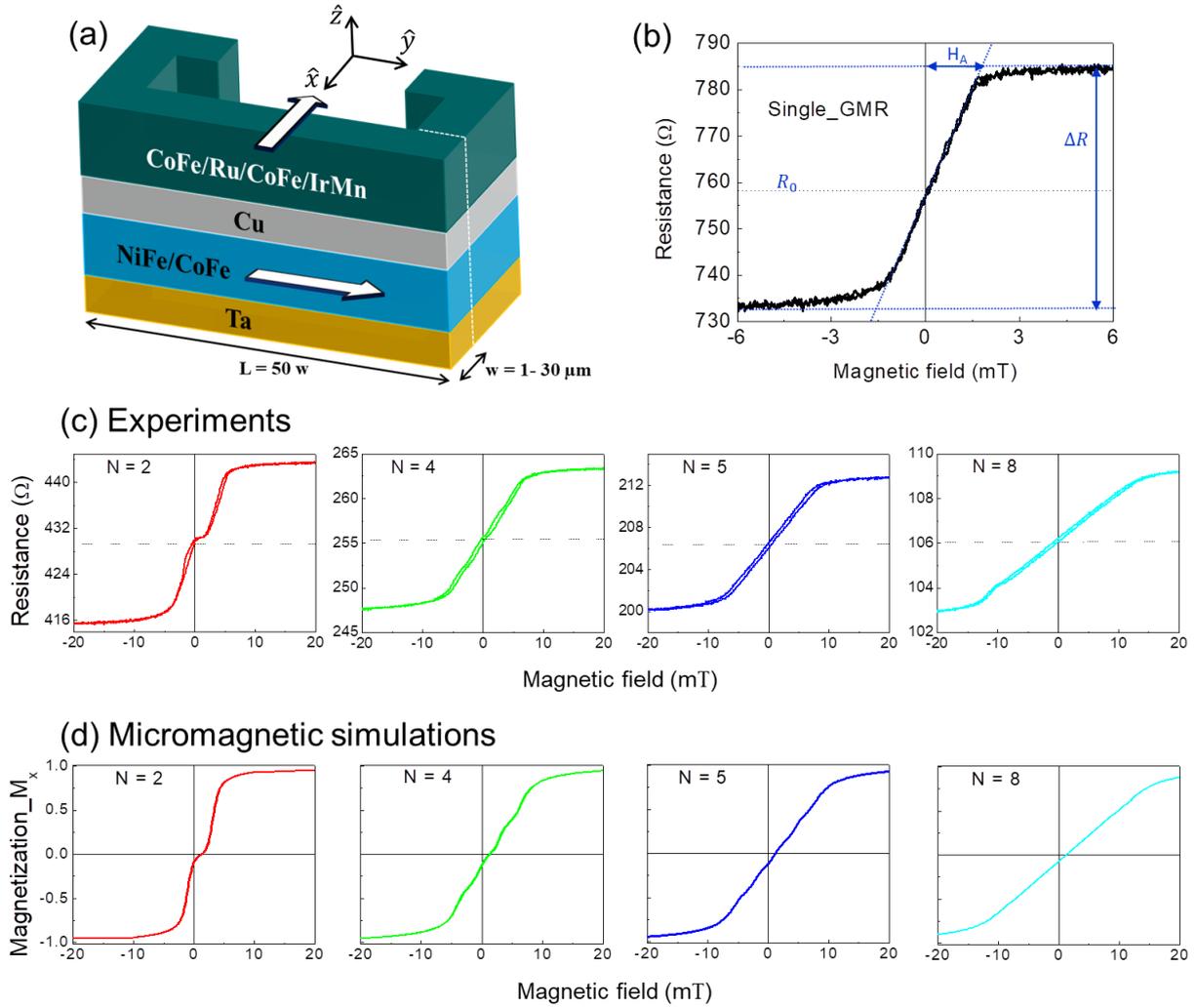

*Figure 2. Magnetoresistance in microdevices: Experiments vs simulations. (a) Scheme of yoke-shaped GMR based on Top-SAF spin valves. (b-c) Experimental RH curves for devices composed by a single GMR device, 2, 4, and 8 GMRs. (d) Magnetization reversal curves computed by OOMMF micromagnetic simulations for N-free layer multi stacks, N=2,4, 5 and 8. The dimension of the free layer is a rectangular prism of 200 x 4 x 0.007 μm³. We assume that NiFe and CoFe layers are perfectly coupled by exchange with the following averaged magnetic parameters: saturation magnetization $M_s$= 850 kA/m and exchange constant $J_s$=10 pJ/m ,weak uniaxial magnetic anisotropy $K_u$=200 J/m³ and Neel coupling distribution ($H_{n1}$=0.8 mT, $H_{n2}$=1.3 mT, $H_{n3}$=1.4 mT, $H_{n4}$=1.5 mT and $H_{n5}$=$H_{n6}$= $H_{n7}$= $H_{n8}$=1.6 mT). The gap (vertical separation) between consecutive free layers is 28 nm.*

The GMR ratio in yoke-shaped sensors follows the same evolution than in thin films: The MR ratio slightly decreases from N=1 to N=12, Figure 3(a). Interestingly, the effective anisotropy field $H_A$ scales as ~N/w in the linear regime, Figure 3(b). This means that the linear magnetic field range can be widely tuned through N in multi-GMR sensors without modifying its area: up to 100 mT for N=12 and w=1 μm. The magnetic anisotropy field evolution can be explained by the dominant shape anisotropy in patterned devices and can be approximated in multi-GMRs as:

$$H_{SH} = \mu_0 M_s \frac{Nt_{FL}}{w} \qquad (1)$$

where $M_s$ is the saturation magnetization of the free layer. We have found a good agreement between the experimental effective anisotropy and Eq. (1), in particular in the narrow width regime (w < 10 µm) where the shape anisotropy is dominant over the bulk magnetocrystalline anisotropy. At higher width, the effective anisotropy will reach a plateau given by the bulk magnetocrystalline anisotropy. Such analytical expression for multi-GMR is equivalent to the shape anisotropy in single GMRs[28] but taking into account the total free layer thickness $Nt_{FL}$. However, compared to single GMRs, it is not possible to tune the magnetic field range in the same way through the NiFe layer thickness. For NiFe layer thicker than 10 nm, the free layer is not uniform and the linearity and GMR ratio are strongly degraded (See Supplementary Figure S2).

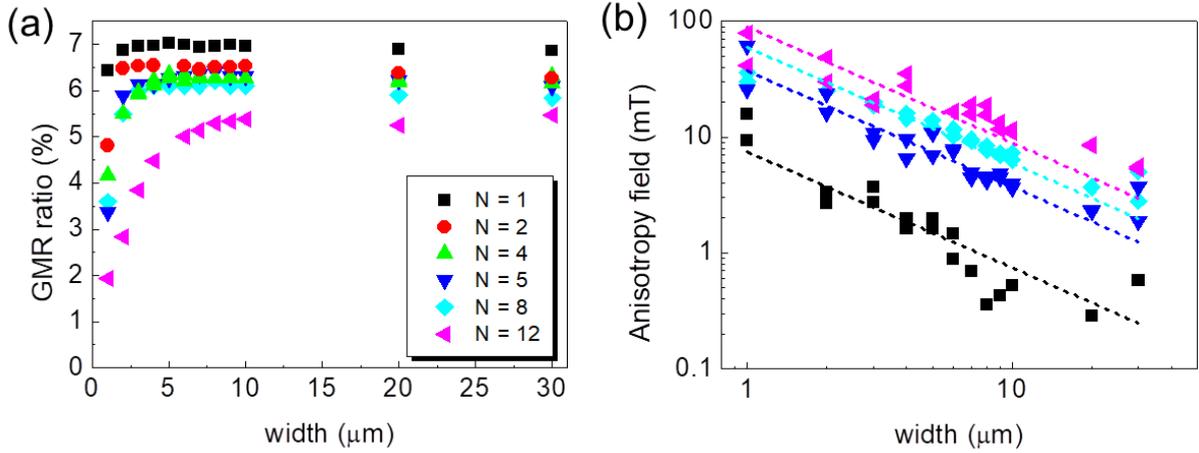

*Figure 3*. *(a) GMR ratio and (b) effective anisotropy field as a function of width for yoke-shaped multi-GMR sensors (N=1-12). The decrease of GMR ratio at low width is artificial and due to the non-saturation in the resistance versus field measurement. The dashed lines correspond to the analytical calculation from Equation (1) using $M_s$= 850 kA/m and $t_{FL}$=7 nm, respectively.*

## B. Micromagnetic simulations

To gain a better understanding about the correlation between the additive dipolar and Neel coupling with the magnetization reversal of multi-GMRs, we have performed full micromagnetic simulations using OOMMF[21]. First, we have computed the hysteresis loops under magnetic field $H_x$ (along the pinned layer magnetization direction) as a function of N and at constant width w= 4 µm, figure 2(d). The hysteresis loop is the total magnetization of the N free layers rotating from P ($\theta = \pi$) to AP ($\theta = 0$) configuration, $M_x = M_s \cos\theta = M_s \frac{\sum_{i=1}^{N} \cos\theta_i}{N}$, under the influence of external magnetic field, MC and shape anisotropies, and the Neel and dipolar couplings. Note that the magnetoresistance of the sensor, Figure (2c), is directly proportional to the $\hat{x}$ component of total magnetization according to $R = R_0 +$ $\Delta R \cos \theta = R_0 + \frac{\Delta R}{M_s} M_x$, where $R_0 = R(H_x = 0)$ and $\Delta R = \frac{R_P + R_{AP}}{2}$. We can observe a very good agreement between the experiments and simulations. In particular, in Figure 2, the simulations reproduced the evolution from step-like to linear behavior around N=5. Note that such transition can slightly vary depending on the value of the uniaxial magnetic anisotropy constant $K_u$ (magnetocrystalline anisotropy). Here the best agreement with experiments has been obtained for $K_u$=200 J/m$^3$ which is coherent with literature[22]: lower $K_u$ values lead to lower threshold N and vice versa (not shown here). Moreover, the simulations also display that the effective anisotropy field scales with N in the linear region in good agreement with Fig. 3(b) and Eq. (1). In the simulations, we have considered the Neel coupling distribution as an individual offset field in each free layer ($H_{n1}$, $H_{n2}$,… $H_{nN}$) taking into account the experimental results in Figure 1.

In order to quantify the role of the two main couplings (additive dipolar vs Neel) in multi-GMRs, we have repeated the same micromagnetic simulations without considering the Neel coupling distribution, that is, setting $H_{n1}$=$H_{n2}$=…=$H_{nN}$=0 (See Supplementary Figure S4). We can observe that the yoke-shaped sensor exhibits very similar response with and without Neel coupling distribution in the linear regime (N ≥ 5), while differences are observed in step-like regime (N < 5). The presence of Neel coupling distribution increases the step length, that is, the separation between the N individual jumps.

To shed more light into the magnetization rotation mechanism in multi-GMRs, we have analyzed the magnetization rotation of each individual free layer. Figure 4(a) displays the reversal magnetization of each free layer for selected number of GMRs (N=2, 4 and 8) with Neel coupling variation. The free layer 1 corresponds to bottom layer with lowest roughness while free layer N is the top layer with highest roughness. At double GMR N=2, in the step-like regime, the magnetization rotation mechanism of the free layers is sequential: the first (second) layer rotates from P ($M_x = -1$) to AP ($M_x = +1$) configuration while the second (first) is blocked in P (AP) state. The sequential mode start to vanish as N increases leading to a more complex rotation scenario at N=4. At higher N=8, in the linear regime, all free layers rotate almost simultaneously in the same magnetic field range towards parallel rotation mode. Interestingly, the magnetic susceptibility of each free layer is strongly reduced in the parallel mode while remains almost unaltered for the sequential mode.

The magnetization reversal of the layers is determined by the competition between the different energies, anisotropies and couplings. In particular, the additive dipolar coupling between the free layers is antiferromagnetic and favors an antiparallel configuration between the free layers, upper Fig. 4(b). The Neel coupling, as it is different for each free layer, breaks the symmetry of the system, lower Fig. 4(b). Due to the Neel coupling, each free layer rotates at a different field. These two couplings, Neel and dipolar, explain the observed behavior in Fig 4 and S4.

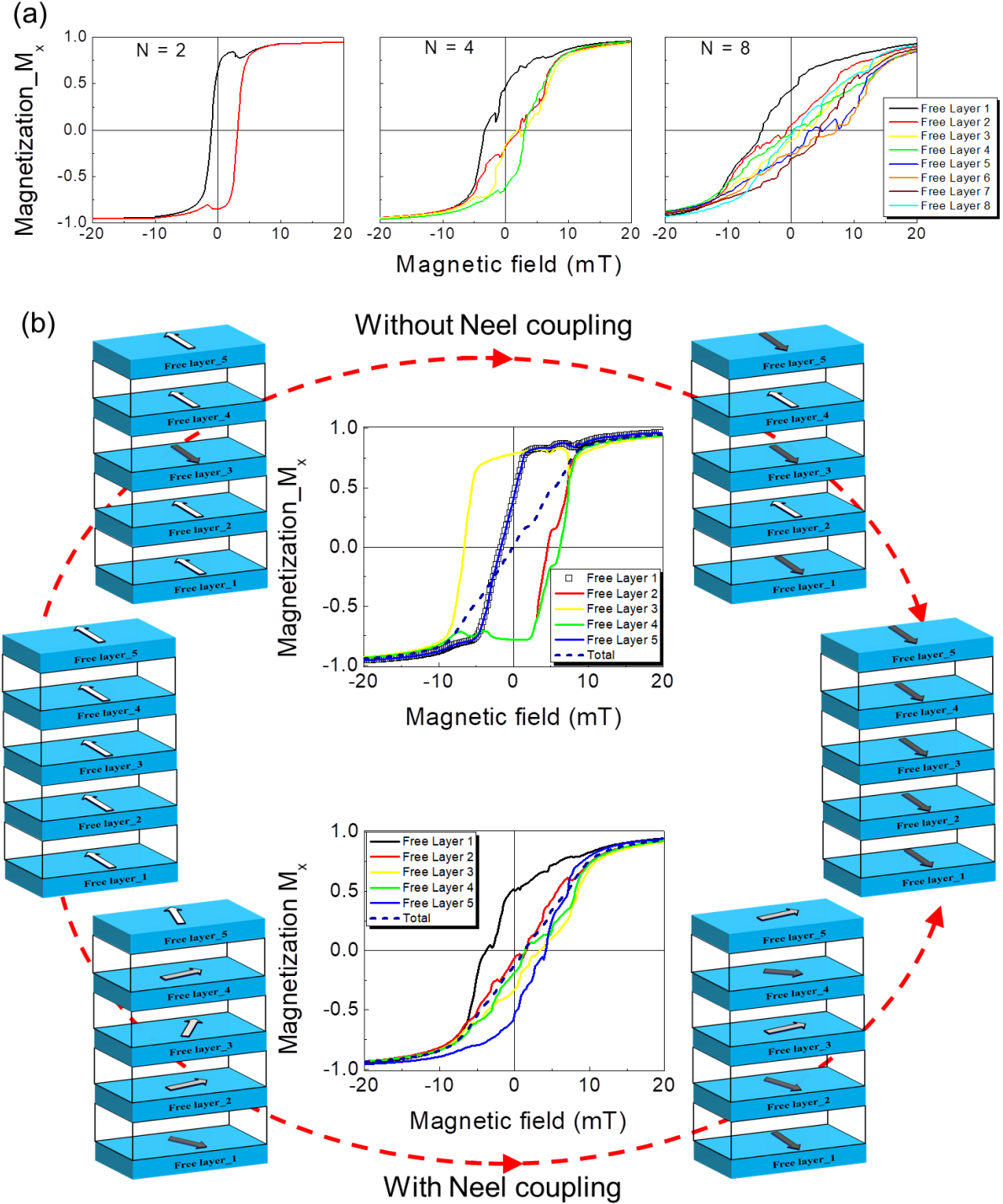

*Figure 4. Magnetization reversal of each individual free layer by OOMMF micromagnetic simulations.* *(a) Multi-GMR devices composed by 2, 4 and 8 GMRs with Neel coupling variation and (b) Microdevice composed by 5 GMRs with and without Neel Coupling variation. The Neel coupling distribution is simulated by different bias fields along the SAF pinned layer direction ($H_{n1}$=0.8 mT, $H_{n2}$=1.3 mT, $H_{n3}$=1.4 mT, $H_{n4}$=1.5 mT and $H_{n5}$=$H_{n6}$= $H_{n7}$= $H_{n8}$=1.6 mT). The dimension of the free layer is a rectangular prism of 200 x 4 x 0.007 $\mu m^3$. We assume that NiFe and CoFe layers are perfectly exchange coupled with the following magnetic parameters: saturation magnetization $M_s$= 850 kA/m and exchange constant $J_s$=10 pJ/m and weak uniaxial magnetic anisotropy $K_u$=200 J/$m^3$. The gap (vertical separation) between consecutive free layers is 28 nm. The external magnetic field along pinned layer direction is swept between ±20 mT.*

For N=2, the Neel coupling favors the step like behavior of the hysteresis cycle. Indeed, the free layer with the smallest Neel coupling rotates first and allows the antiferromagnetic configuration to be more stable than without the Neel coupling. When N increases, a complex reversal of the layers, forming a spin-canted antiferromagnetic helix, appears. Due to the Neel coupling, the free layer 1, at the bottom rotates first. Due to antiferromagnetic dipolar coupling, a "spin-flop" rotation of the other free layers appears and creates step by step a complex antiferromagnetic helix rotation, lower Fig. 4(b). Without Neel coupling, the symmetry of the system is conserved and the free layers rotate in a more sequential mode, upper Fig. 4(b).

In conclusion, the presence of Neel and dipolar couplings induces complex reversal behaviour of the free layers. A low N it favours a step like resistance behaviour while at higher N, a linearized multi-GMR response is observed with a spin-canted antiferromagnetic helix rotation.

## V. Multi-GMR sensor performance

In the last section, we have characterized the multi-GMR sensor performance from noise measurements, Figure 5, and extracted the main parameters: sensitivity ($s$), noise ($S_V$) and detectivity ($D = \frac{S_V}{sV_{in}}$). The evolution of these sensor parameters with the number of GMR repetitions N, width and input voltage ($V_{in}$) have been correlated to the additive dipolar coupling mainly characterized by N and w. For simplicity, experimental data for multi-GMRs N=2 and N=5 have been not included in Figure 6. Double-GMRs exhibit a strong step-like behavior and remarkable RTN noise in many devices and therefore sensitivity and noise evolution differ from the general trend. And multi-GMRs with N=5 exhibit very similar values than N=4. Such additional data can be found in Supplementary Figure S5.

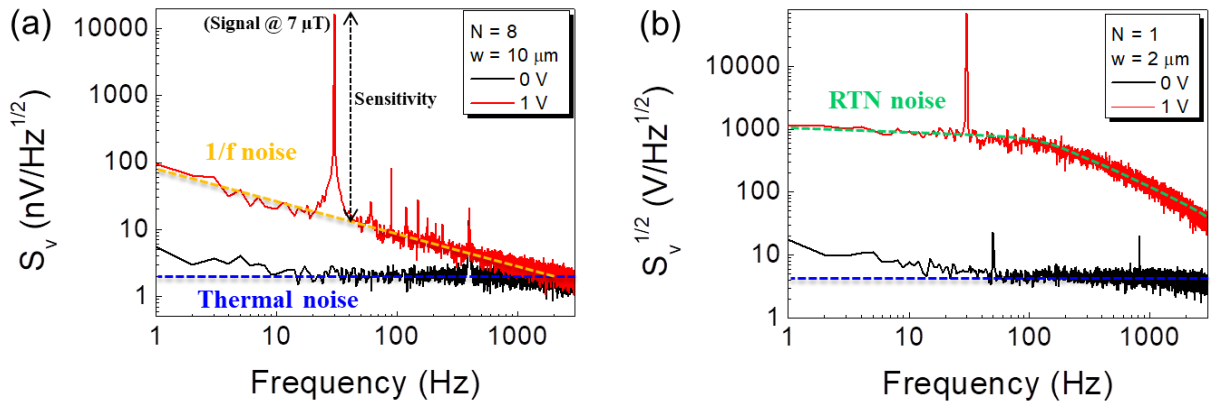

*Figure 5. Noise spectra in multi-GMR devices.* Noise spectra of a multi-GMR sensor (N = 8 & w = 10 μm) and (b) single GMR (N = 1 & w = 2 μm) consisting of different sources: 1/f, thermal and RTN noise.

## A. Sensitivity

The sensitivity is directly extracted from the reference AC magnetic signal at 30 Hz which is basically the voltage amplitude divided by the equivalent field $H_{eq}$ = 7 µT and input voltage $V_{in}$=1V. The sensitivity of multi-GMRs as a function of width for selected N = 1, 4, 8 and 12 is displayed in Figure 6(a). The sensitivity scales with the width when the shape anisotropy is dominant mechanism in the narrow range and it is likely constant in the wide range where the volume magnetocrystalline anisotropy overcomes the shape anisotropy. Interestingly, the linear evolution of sensitivity expands to wider range as N increases: up to w=4, 8 and 15 µm for N=4, 8 and 12, respectively. As expected, the dipolar coupling is strongly correlated to the shape anisotropy because both are determined by the same parameters (width, thickness and saturation magnetization of the free layer). Therefore the number of GMR repetitions N plays a similar role than the thickness of the free layer and the sensitivity in the strong dipolar regime (narrow range) decays roughly as $\sim \frac{1}{N}$ (see dashed lines in Figure 6(a)) according to the following expression:

$$s_{Dip} = \left(\frac{GMR}{2\mu_0 M_s t}\right)\frac{w}{N} \tag{2}$$

Concerning the saturation range at wider width, the asymptotic sensitivity $s_{MC}$ is very different between single GMR and multi-GMR. It might be ascribed to a change of effective magnetic (magnetocrystalline) anisotropy as a consequence of roughness propagation from the first to N-th free layer.

Finally, the sensitivity in the single GMR decays strongly for wider devices (w ≥ 10 µm) because the offset field is close to the anisotropy field (Magnetocrystalline contribution) and such devices operate close to the saturation regime similarly to the thin film stack in right-top panel in Figure 1(b). In multi-GMR stacks, the additive dipolar coupling extends the range of maximum sensitivity around zero field (linearity) to wider range but the same decays should be observed for wider devices than the maximum width analyzed in this study (w > 30 µm).

We should point out that the sensitivity can be also calculated from magnetotransport measurements in Figure 2(b-c) when the RH loops are practically anhysteretic. This is valid for narrow devices, where the sensitivity is just the resistance slope around zero magnetic field (±0.2 mT). As the hysteresis becomes noticeable at critical width, the sensitivity from magnetotransport deviates from those extracted from noise measurements and minor loops from irreversible mechanisms should be taking into account for the correct sensitivity estimation.

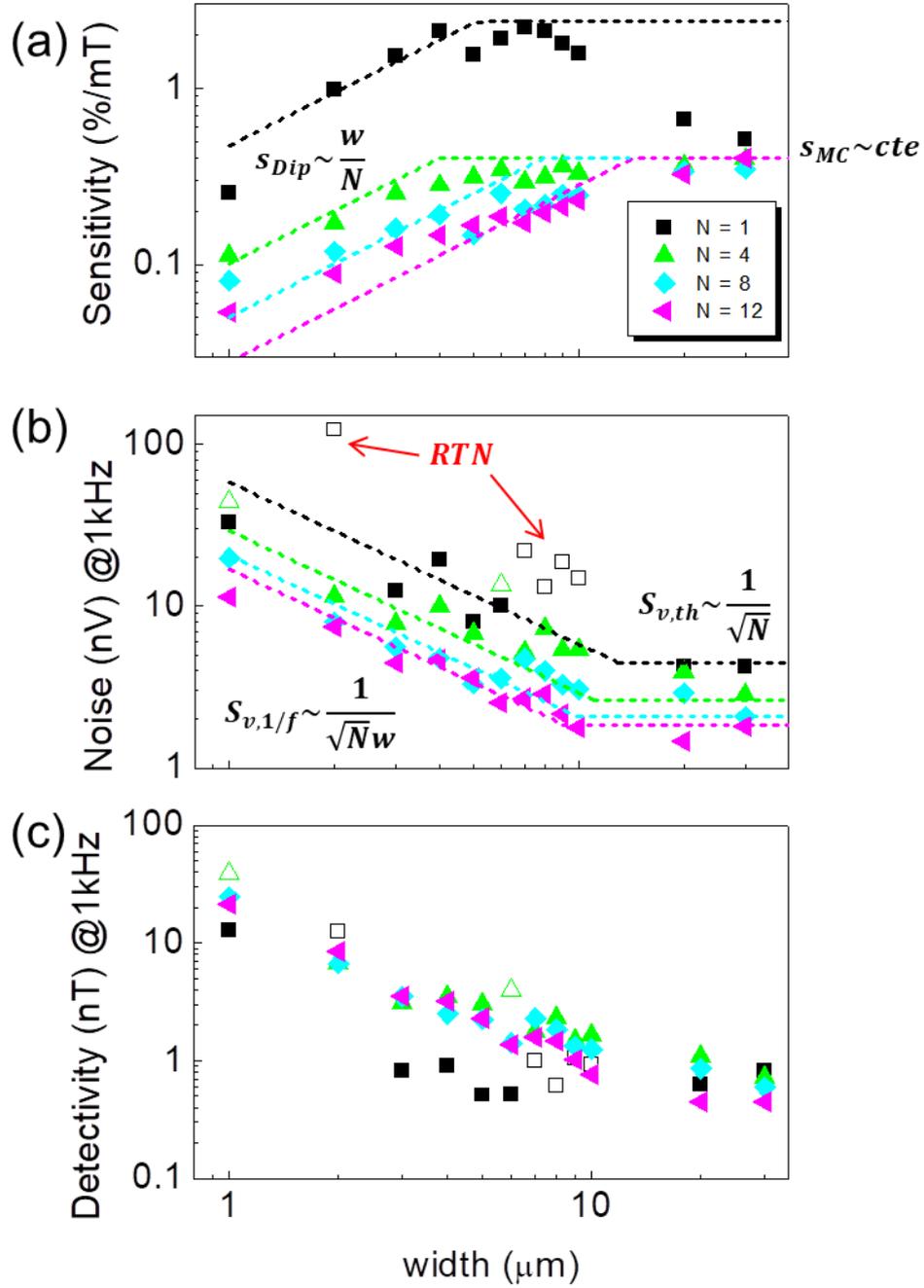

*Figure 6. Noise measurements in multi-GMR devices as a function of the width for constant input voltage ($V_{in}=1V$).* (a) Sensitivity, (b) noise (1 kHz) and (c) detectivity ( 1kHz), as a function of the sensor width (w = 1-30 µm) for selected GMR numbers. Open symbols in (b) and (c) represents devices with large magnetic noise (1/f or RTN). Dashed lines in (a) correspond to theoretical Eq. (2) with $M_s$= 850 kA/m, t=7 nm and averaged GMR ratio for each multi-GMR system (7% for N=1, 6% for N=4 and 8 and 5% for N=12) according to Figure 3(a). Asymptotic sensitivities $s_{MC}$ are manually added in (a) for better visualization. Dashed lines in (b) corresponds to theoretical fit from Eq.(3) in the 1/f regime and Eq.(4)in the thermal regime using average $\alpha = 1.2 \times 10^{-12} \mu m^3$ for all devices, single GMR resistance $R_1 = 800\Omega$ and temperature T=300 K. A vertical offset of 0.8 nV has been added in the thermal regime for taking into account the noise floor of the experimental set-up.

## B. Noise

The power spectral density $S_V$, Figures 5(a-b), associated to the noise in GMR sensors can exhibit different origins[29]: 1/f ($S_{V,1/f}$) and thermal noise ($S_{V,th}$) and in some cases undesirable Lorentzian random telegraphic noise, RTN ($S_{V,RTN}$). In absence of RTN, the noise is dominated by 1/f noise at low frequencies and thermal noise at high frequencies, Figure 5(a). In particular, for a multi-GMR with N=8 and w=10 μm, the 1/f noise is found below 2 kHz and the thermal noise above this corner frequency. The noise level at 1 kHz of multi-GMR sensors (N=1, 4, 8 and 12) as a function of the width and for a constant input voltage $V_{in}$= 1V is shown in Figure 6(b). For w ≤ 10 μm, the sensors are in the 1/f regime and the noise level decreases with the magnetic volume of the free magnetic layer according to Equation (3):

$$S_{v,1/f} = \sqrt{\frac{\alpha}{V_{FL}f}} V_{in} = \sqrt{\frac{\alpha}{50Ntf}} \frac{V_{in}}{w} \quad (3)$$

where $\alpha$ is the Hooge constant and it can have electric and magnetic origins. It is very relevant parameter when comparing the performance of different magnetic systems. The yoke-shaped sensors have the same aspect ratio, L = 50 w, and the total magnetic thickness is $N \times t_{FL}$, then the noise in Figure 6(b) decreases in the 1/f regime as $\sim \frac{1}{\sqrt{Nw}}$ in reasonable good agreement with Eq. (3) using the same Hooge constant $\alpha = 1.2 \times 10^{-12} \mu m^3$ for all devices. Nevertheless, some devices deviate from the Eq. (2) due to the presence of large magnetic noise at low frequency (1/f or RTN) induced by magnetic fluctuations in the free layer(s) and therefore increasing drastically $\alpha$ (the magnetic origin). The magnetic noise (1/f or RTN) is noticeable in single GMR with very weak shape anisotropy (w = 6-10 μm) and multi-GMRs with weak dipolar coupling N≤4. Magnetic noise is rarely observed in narrow devices with strong shape anisotropy (single GMR, w= 2 μm, Figure 5(b)) which might be due to some defects created during the microfabrication process (in particular during the etching process) that might act as a seed of reversed domains nucleation[30]. However, for multi-GMRs with N>4, the magnetic noise and RTN is not observed in the full width and voltage range (up to 3V). An explanation is that the increased dipolar coupling stabilizes the magnetic domains and suppresses the magnetic noise.

For w > 10 μm, the sensors reach the thermal regime where the noise only depends on the total resistance of the devices according to:

$$S_{v,th} = \sqrt{4k_B T R_N} = \sqrt{\frac{4k_B T R_1}{N}} \quad (4)$$

Each GMR repetition reduces the total resistance according to $R_N=R_1/N$. Therefore in the thermal noise in Figure 6(b) is roughly reduced as $\sim \frac{1}{\sqrt{N}}$. Note that Eq. (4) provides lower noise levels than experiments as the current and voltage input noise of the preamplifier have to be added quadratically.

## C. Detectivity

Finally, the magnetic field detectivity $D$ of multi-GMR sensors has been evaluated as a function of width (Figure 6(c)) and input voltage (Figures 7). The detectivity evolution at 1 kHz with the width in Figure 6(c) reflects the complex competition between sensitivity (both components, dipolar $s_{Dip}$ and magnetocrystalline $s_{MC}$) and noise (both regimes, 1/f and thermal) according to $D = \frac{S_V}{sV_{in}}$. Indeed, a multi-GMR could be in four regimes. At low width and low frequencies, the dipolar coupling and the 1/f noise are dominating the sensitivity and the noise behavior respectively. The detectivity (without considering RTN) scales as $\sim \frac{\sqrt{N}}{w^2}$ according to:

$$D_{1/f}^{Dip} = \frac{S_{v,1/f}}{s_{Dip}V_{in}} = \left(\frac{2\mu_0 M_s \sqrt{\alpha t}}{GMR\sqrt{50f}}\right)\frac{\sqrt{N}}{w^2} \quad (5)$$

At higher width and low frequencies, the magnetocrystalline anisotropy overcomes the shape anisotropy, $s_{MC}$ should be considered instead of $s_{Dip}$, and the detectivity decays slower as $\sim \frac{1}{w}$ according to:

$$D_{1/f}^{MC} = \frac{S_{v,1/f}}{s_{MC}V_{in}} = \sqrt{\frac{\alpha}{50Ntf}}\frac{1}{s_{MC}w} \quad (6)$$

At higher width (w ≥ 10 µm) and high frequencies (in the thermal regime), the dipolar coupling is negligible, magnetocrystalline anisotropy dominates and sensitivity converges to a constant value $s_{MC}$. Therefore the detectivity in the thermal regime is independent of the width scaling as $\frac{1}{\sqrt{N}}$ according to:

$$D_{th} = \frac{S_{v,th}}{s_{MC}V_{in}} = \sqrt{\frac{4k_B T R_1}{N}}\frac{1}{s_{MC}V_{in}} \quad (7)$$

The last regime, low width and high frequencies, is not investigated in this paper since the frequency (~MHz) has to be very high to achieve the thermal regime.

In general, the single GMR exhibits a better detectivity in the narrow width range (w ≤ 8 µm) while multi-GMR performance is more competitive in the large width range (w ≥ 10 µm) in good qualitatively agreement with Eqs. (5-7). The 1/f detectivity variation for single GMRs does not follow the Equation (6) due to presence of remarkable RTN in the intermediate width range (w = 7 – 10 µm). Despite the remarkable magnetic noise and RTN of the single GMR, its much higher asymptotic sensitivity $s_{MC}$ still leads to lower or similar detectivities levels than the multi-GMR sensor in the intermediate width range.

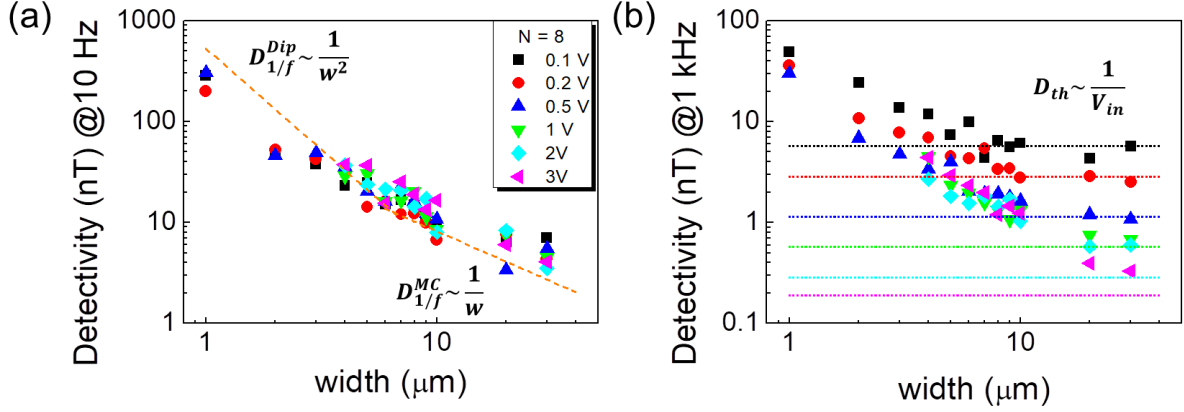

*Figure 7. Detectivity evolution with the input voltage. Detectivity as a function of sensor width for selected input voltages range (0.1 – 3V) for a sensor composed by 8 GMRs for two different frequencies: (a) 10 Hz and (b) 1kHz. Dashed guidelines are manually added for better visualization. Dashed lines in (a) corresponds to theoretical fits from Eq.(5) for= $2 \times 10^{-12} \mu m^3$ in the narrow range (averaged GMR=6 %, $M_s$= 850 kA/m and t=7 nm) and Eq.(6) for= $3 \times 10^{-12} \mu m^3$ in the wide range (asymptotic sensitivity $s_{MC}$= 0.4 %/mT), respectively. Dashed lines in (b) corresponds to theoretical fit from Eq.(7) using single GMR resistance $R_1 = 800\Omega$ and temperature T=300 K. A residual noise floor of 1 nV has been taken into account in the thermal regime.*

For input voltage above 1V, the detectivity of single GMRs is strongly degraded by the presence of remarkable RTN. The multi-GMR is more robust against magnetic noise and RTN is not observed for N>4 (in the full wide range) up to maximum input voltage 3V as shown Fig. 7. Above 3V, the high current density heats the stacks leading to a progressively decrease of the sensitivity (not shown here). The detectivity for a multi-GMR with N=8 has been analyzed in an input voltage range between 0.1 and 3 V, Figs 7. In the 1/f noise regime at 10 Hz, figure 7(a), the detectivity is constant with the input voltage and exhibits two different contributions ($D_{1/f}^{Dip}$ and $D_{1/f}^{MC}$), according to Eqs (5) and (6). Good agreement between experiments and theory in Fig. 7(a) has been obtained for $\alpha = 2 \times 10^{-12} \mu m^3$ in the dipolar regime and $\alpha = 3 \times 10^{-12} \mu m^3$ in the magnetocrystalline regime. This result suggests that $\alpha$ can also vary between devices with different widths in the same multi-GMR systems being a possible source of discrepancies between theory and experiments in Fig. 6(b) and 7(a). In particular, the hysteresis is more relevant in wider devices leading in general to higher magnetic noise and therefore higher $\alpha$.

Finally the detectivity is reduced roughly as $\frac{1}{V_{in}}$ in the thermal regime at 1 kHz, Fig. 7(b), in good agreement with Eq. (7). Note that the threshold width for reaching the thermal regime depends on the voltage and higher voltages shift the 1/f – thermal transition to wider widths: from $w_c$~6 μm to ~20 μm when voltage increases from 0.1 to 3 V, respectively. Similarly to Fig. 6(b), additional residual noise ~$1\ nV$ is needed to observe good quantitative agreement between experiments and theory. Discrepancies at maximum voltages might be ascribed to progressive heating of the multi-GMR stacks leading to progressive reduction of sensitivity and/or increase of thermal noise (through the temperature).

# VI. Conclusion

To conclude, all important considerations on multi-GMR sensors have been highlighted in Table I. In summary, we have fabricated and characterized (from magnetotransport and noise measurements) for the first time multi-GMR sensors based on N (up to 12) spin valves vertically stacked which keep good GMR ratio, linearity and low roughness propagation. Combining magnetoresistance measurements with micromagnetic simulations, we have identified the two main magnetic mechanisms: Neel coupling distribution induced by the roughness propagation and additive dipolar coupling between the free layers. From experiments, we have found a smooth crossover from step-like to linear behaviour at N=5 (w=4 µm) in excellent agreement with micromagnetic simulations. At low N, a step-like response is stabilized by the combination of both additive dipolar and Neel couplings where the reversal magnetization of the N dipolarly-coupled free layers is likely sequential maintaining the antiferromagnetic order at zero field. As N increases, the Neel coupling variation between the free layers induces a more complex reversal in a spin-canted antiferromagnetic helix, resulting in a linearization of the resistance response.

|  | **Multi-GMR based on N spin valves** | **Single GMR** |
|---|---|---|
| **GMR ratio** | Smooth decrease from 7% to 6% for max N=12 | 7% |
| **Roughness (Neel coupling)** | Low roughness propagation up to N=12 (Max offset ~ 3 mT at N=12) | Offset < 0.8 mT |
| **Linearity** | tuned by N in a large magnetic field range (up to ±100 mT for N=12 and w=1µm) | Limited to few mT (±10 mT for w=1µm) |
| **Sensitivity** | Decreases as $\sim \frac{1}{N}$ (strong dipolar regime) | max $s \sim 2\ \%/mT$ (w= 4-10 µm) |
|  | Converges to $s \sim 0.4\ \%/mT$ (weak dipolar regime) |  |
| **Noise** | Decreases as $\sim \frac{1}{\sqrt{N}}$ in the 1/f and thermal regime. Robust against RTN for N>4 up to 3V in full width range | stronger magnetic noise (w= 6-10 µm & $V_{in} \geq 1V$) |
| **Detectivity** | higher in the 1/f, $D \sim \sqrt{N}$ | Optimal D for $V_{in} \leq 1V$ |
|  | Lower in the thermal regime at high $V_{in}$ |  |

*Table I. Main relevant points of Multi-GMR sensor performance compared to single GMR in terms of GMR ratio, roughness, linearity, sensitivity, noise and detectivity.*

Compared to single GMRs, multi-GMR sensor performance exhibits very good linearity in a much larger magnetic field range (one order of magnitude larger, up to ±100 mT) without remarkable degradation of GMR ratio, with noise reduction ($\sim \frac{1}{\sqrt{N}}$) in both 1/f (via magnetic volume) and thermal (through the total resistance) regimes and with more robustness against magnetic noise (1/f and/or RTN). However, the sensitivity roughly decreases as $\sim \frac{1}{N}$ due to the strong dipolar coupling between neighboring free layers and it has a negative impact on the detectivity at 1/f noise regime. Therefore, lower detectivities in multi-GMRs are only achieved in the thermal regime for the widest devices with negligible dipolar coupling. Finally, the analytical equations predict reasonably well the experimental evolution of sensitivity and noise with number of GMR repetitions, width and input voltage and can be used as general guidelines to design multi-GMR sensors with optimized performance.

This pioneer result in multi-GMR stack opens the route towards spintronic sensors coupled in 3D with compact footprint and additional tunable parameter through the number of vertical stacked GMRs (N). The next generation of multi-GMR sensors with improved detectivity will need to face some fabrication challenges in order to minimize the sensitivity loss (induced by dipolar coupling): increasing the spacer thickness and number of GMRs avoiding a strong roughness degradation and a very long fabrication process and/or growing multi-GMR stacks based on synthetic antiferromagnet free layers[31] without a remarkable degradation of GMR ratio, linearity and offset.

## Acknowledgement

This research has been supported by the *Agence Nationale de Recherche* (France) through grant n°ANR-17-CE19-0021-01 (NeuroTMR) and the *Commissariat à l'énergie atomique et aux énergies alternatives* (CEA) for the internal funded project CARAMEL. AD acknowledges a SNSF mobility fellowship (177732).

# VII. Supplementary information

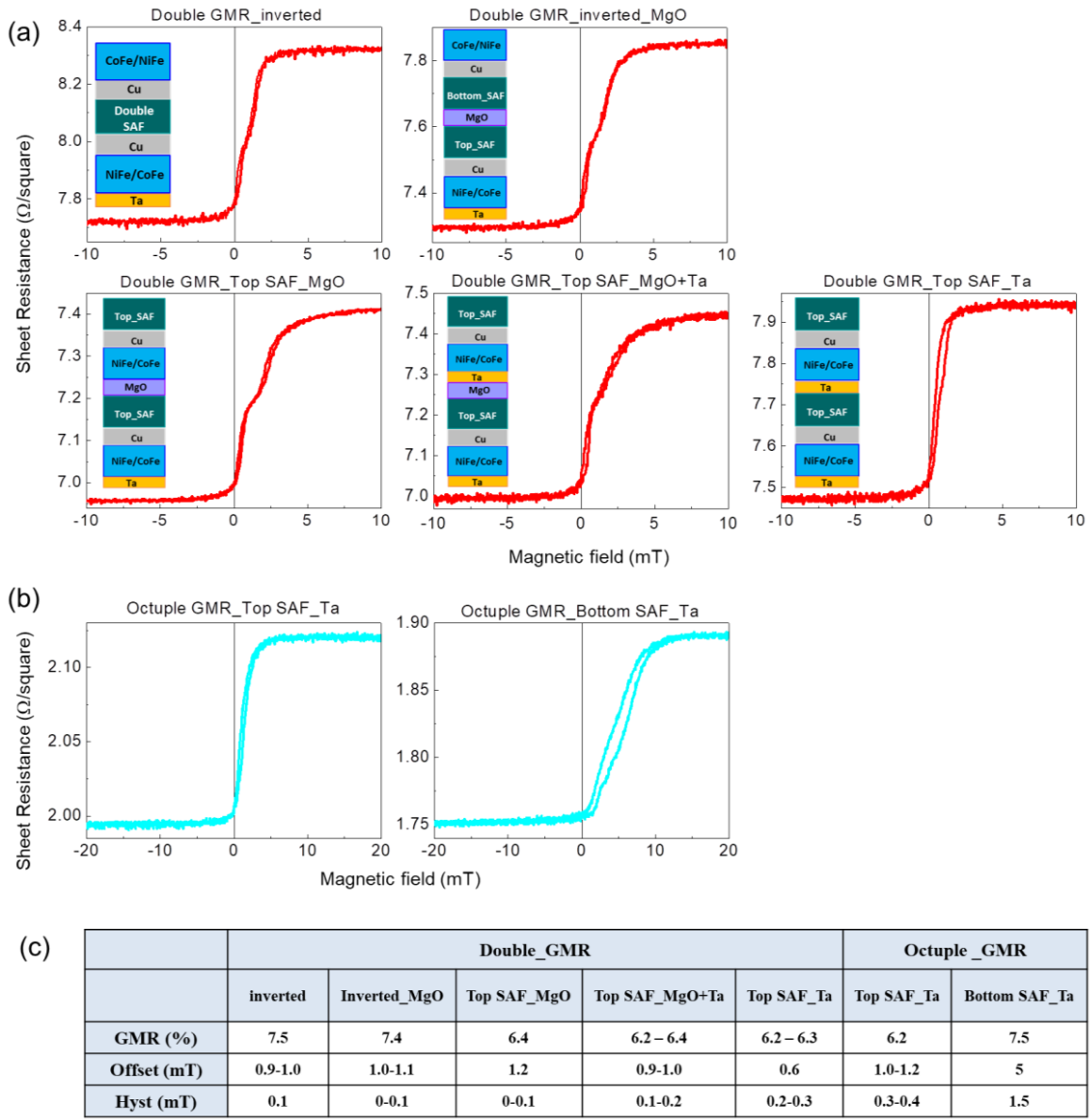

*Figure S1. Different multi-GMR combinations. (a) Double GMRs: inverted (Bottom SAF / Top SAF), inverted with MgO spacer (Bottom SAF / MgO / Top SAF), Top SAF with MgO spacer, Top SAF with MgO+Ta spacer and Top SAF with Ta spacer. (b) Octuple GMRs: Top SAF and Bottom SAF with Ta spacer. (c) Comparative table of GMR ratio, offset and hysteresis for the different multi-GMR systems. All results correspond to film set A.*

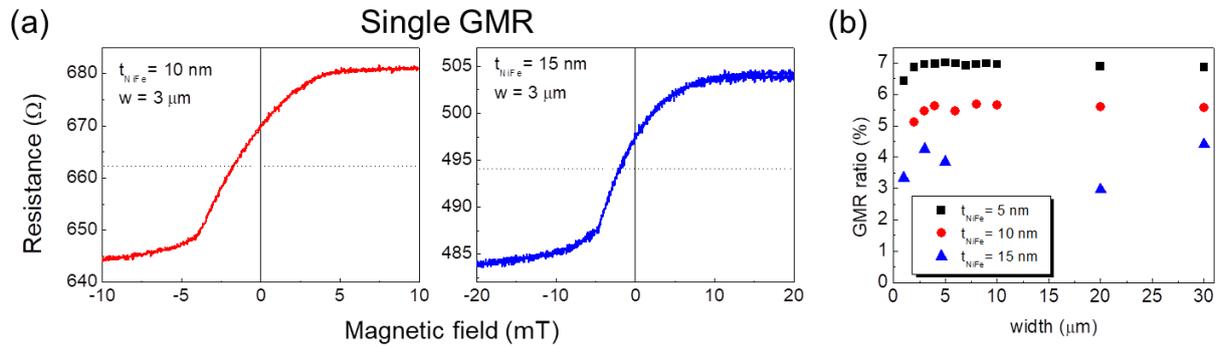

*Figure S2. Evolution Single GMR with NiFe thickness.* (a) Experimental RH curves for single GMR (w=3μm) and (b) GMR ratio as a function of sensor width for selected NiFe thicknesses (5, 10 and 15 nm).

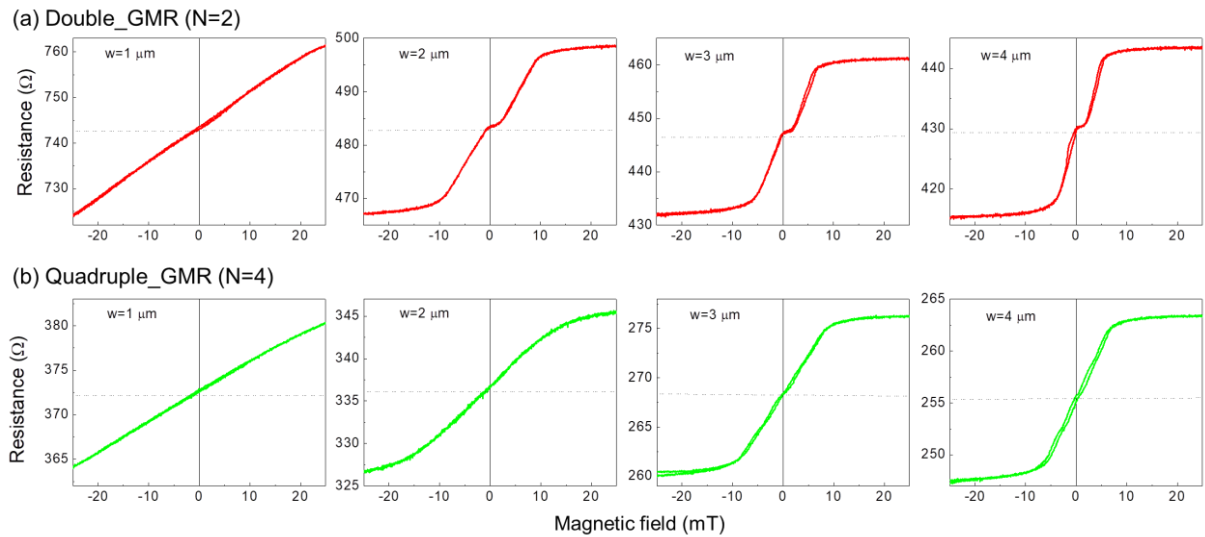

*Figure S3. Magnetoresistance in yoke-shaped multi-GMR sensor.* Experimental RH curves for (a) double and (b) quadruple GMR sensors for different widths w = 1-4 μm.

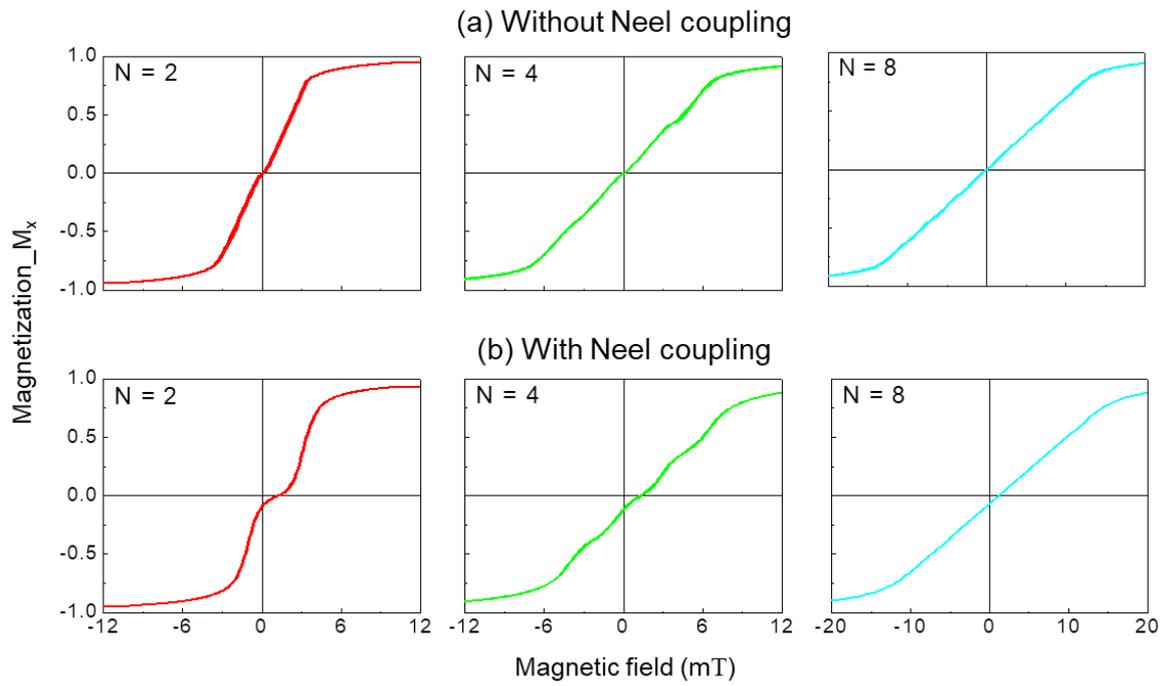

*Figure S4. Role of Neel coupling distribution in micromagnetic simulations.* *Magnetization reversal curves computed by OOMMF for multi-GMR sensors with N=2, 4 and 8 and w=4μm: (a) without and (b) with Neel coupling distribution ($H_{n1}$=0.8 mT, $H_{n2}$=1.3 mT, $H_{n3}$=1.4 mT, $H_{n4}$=1.5 mT and $H_{n5}$=$H_{n6}$= $H_{n7}$= $H_{n8}$=1.6 mT)*

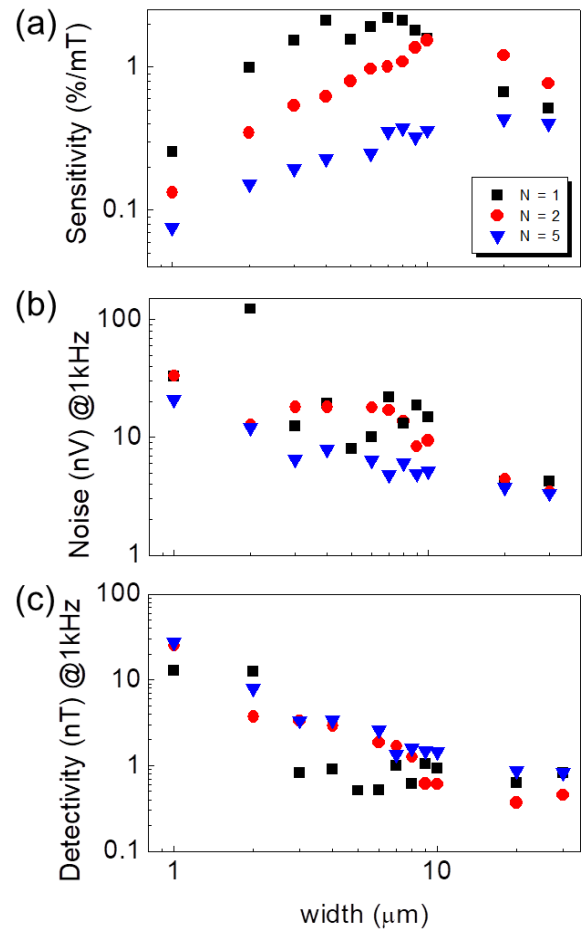

*Figure S5. Noise measurements in multi-GMR devices with N=2 and N=4.* *(a) Sensitivity, (c) noise (1 kHz) and (d) detectivity ( 1kHz), as a function of the sensor width (w = 1-30 μm).*